\documentclass[10pt,pre,a4paper,twocolumn,superscriptaddress,showpacs]{revtex4-1}

\usepackage[utf8]{inputenc}
\usepackage[colorlinks=true]{hyperref}
\usepackage{amsmath,amssymb}
\usepackage{amsfonts}
\usepackage{graphicx}
\usepackage{grffile}%to compile figures with dots in fignames
\usepackage{color}
\usepackage[dvipsnames]{xcolor}

\begin{document}

\title{ Rheology in dense assemblies of spherocylinders: frictional
  vs. frictionless}
\author{Trisha Nath} \affiliation{Institute for Theoretical Physics,
  Georg-August University of G\"ottingen, Friedrich-Hund Platz 1,
  37077 G\"ottingen, Germany} \author{Claus Heussinger}
\affiliation{Institute for Theoretical Physics, Georg-August
  University of G\"ottingen, Friedrich-Hund Platz 1, 37077
  G\"ottingen, Germany}

%\author{Trisha Nath \and Claus Heussinger}
%\institute{Institute for Theoretical Physics, Georg-August
%  University of G\"ottingen, Friedrich-Hund Platz 1, 37077
%  G\"ottingen, Germany}

\begin{abstract}

  Using molecular dynamics simulations, we study the steady shear flow
  of dense assemblies of anisotropic spherocylindrical particles of
  varying aspect ratios. Comparing frictionless and frictional
  particles we discuss the specific role of frictional inter-particle
  forces for the rheological properties of the system. In the
  frictional system we evidence a shear-thickening regime, similar to
  that for spherical particles. Furthermore, friction suppresses
  alignment of the spherocylinders along the flow direction. Finally,
  the jamming density in frictional systems is rather insensitive to
  variations in aspect-ratio, quite contrary to what is known from
  frictionless systems.

\end{abstract}

\maketitle
\section{Introduction}

Flow and arrest in dense dispersions and granular systems is an
important current theme in both materials science and fundamental
research. Computer simulations have been developed to better
understand the very broad range of observed rheological phenomena. One
such phenomenon is discontinuous shear thickening, which signals the
discontinuous increase of flow resistance (viscosity) when the forcing
is increased by only small amounts.  Such a behavior is highly
relevant for many industrial operations, like the mixing or pumping of
cementitious paste or concrete~\cite{FEYS2009510}. Recent work
hints at a mechanism that relies on the direct particle frictional
interactions upon
contact~\cite{0034-4885-77-4-046602,PhysRevE.88.050201,PhysRevLett.111.218301,wyart_cates,comtet,Clavaud5147,PhysRevLett.119.158001}.

Given the relevance of particle rotations for the frictional forces
and the coupling to translational motion~\cite{maiti_claus} we expect
particle shape to be a determining factor for the rheological
behavior, in particular for shear-thickening. Thus, in this work we
study the flow of dense assemblies of frictional particles with
non-spherical shape. Particle shape has been investigated in
connection with the jamming transition in several
publications~\cite{PhysRevE.67.051301,PhysRevLett.118.068002,PhysRevE.97.012905,PhysRevE.97.012909}.

In granular systems numerous parameters such as aspect ratio,
convexity, angularity play important roles to design transport,
conduction, diffusive properties of a material~\cite{C3SM50298H}.  One
of the fundamental observation in systems of non-spherical elongated
particles is the orientation in shear flow, the angle between flow
direction and average orientation of the particles being
nonzero~\cite{campbell_ellipse,guo_wassgren2012,guo_wassgren_2013,PhysRevLett.108.228302,boton}.
Alignment decreases with aspect ratio of the particle, but has zero or
no dependence on strain
rate~\cite{PhysRevLett.108.228302,PhysRevE.86.051304}.  The nematic
order parameter is either monotonic~\cite{reddy_2009,somfai}, or
nonmonotonic function of density depending on the particle shape or
aspect ratio~\cite{2018teitel}. Here, we study the interplay of
rheological properties and shear-induced alignment for model
spherocylindrical particles. With an interest in the shear-thickening
response we aim at comparing frictionless and frictional particles.

\begin{figure*}
  \includegraphics[width=0.29\textwidth]{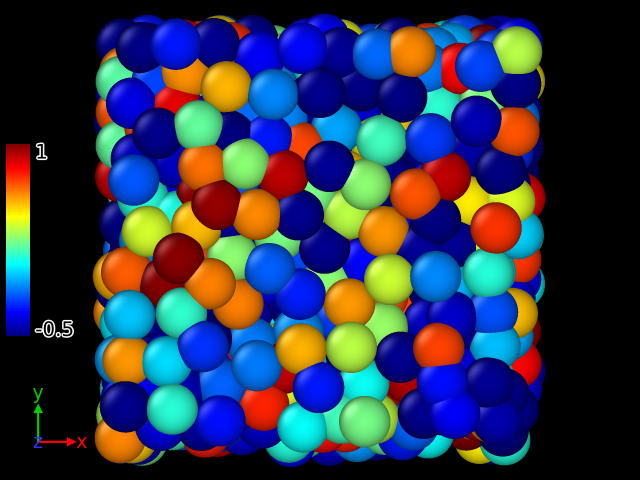}
  \includegraphics[width=0.29\textwidth]{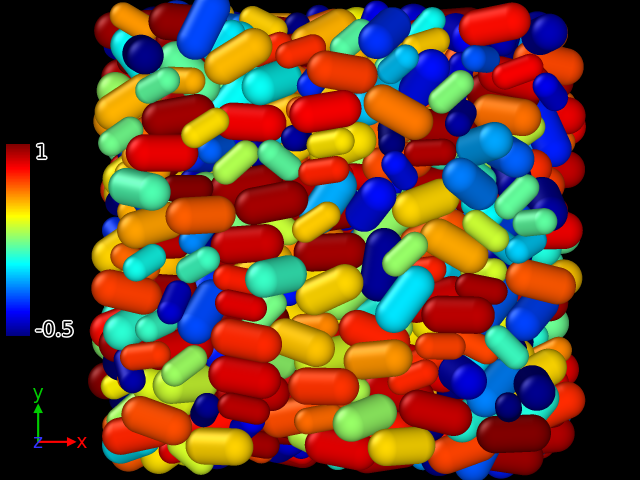}
  \includegraphics[width=0.29\textwidth]{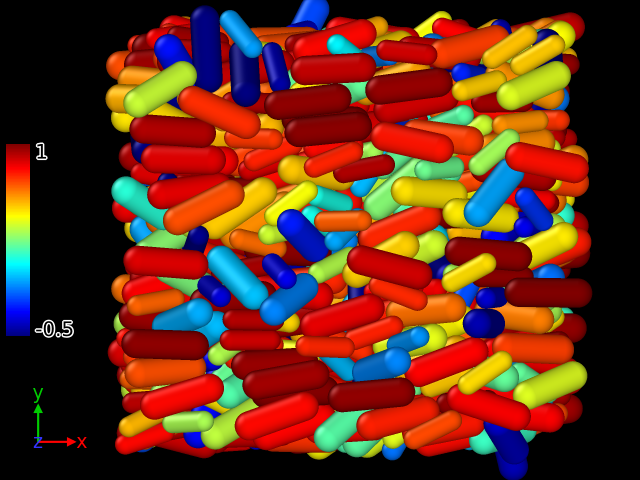}
  \caption{Snapshots of the system at steady state. The particles are
    colored according to the amount of their alignment with the flow
    (Eq.~(\ref{eq:local_align})). The strainrate is $10^{-4}$, with
    particles having (left) $\mathrm{AR}$=0.001, $\phi$=0.582, (middle)
    $\mathrm{AR}$=1.0, $\phi$=0.576 and (right) $\mathrm{AR}$=2.0, $\phi$=0.542. The
    densities are chosen to correspond to the maximum value of the
    global orientational order parameter $\mathbf{S2}$ at the smallest
    available strainrate (see Fig.~\ref{fig:nop}).}
  \label{fig:snapshot_orient}
\end{figure*}

\section{Model}\label{sec:model}
\begin{figure}[b!]
  \includegraphics[width=0.2\textwidth]{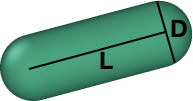}
  \caption{A spherocylindrical particle of aspect ratio $\mathrm{AR}=L/D=2$,
    where $L$ is the length of the ``spine'', i.e. the line segment at
    the center of the cylinder.}
  \label{fig:sphero}
\end{figure}

We perform computer simlations on a 3D system of spherocylindrical
particles using interaction forces %molecular dynamics methods
developed e.g. in Ref.~\cite{Pournin2005}.  A spherocylinder in 3D
consists of a cylindrical part of length $L$ and two hemispherical
caps of diameter $D$ at two ends (see Fig.~\ref{fig:sphero}).  The
spherocylinders are parametrized by their length-to-diameter (aspect)
ratio $\mathrm{AR}=L/D$. Spheres correspond to $\mathrm{AR}=0$.
The system has equal numbers of big and small particles with the same
aspect ratio $\mathrm{AR}$ and a diameter ratio of 1.4. We have
simulated systems of particles with $\mathrm{AR}$=0.001, 0.1, 0.5,
1.0, 1.5 and 2.0, where the particles with $\mathrm{AR}$=0.001 could
be considered as spherical particles.

Homogeneous shear flow is implemented with periodic boundary
conditions in the flow-direction ($\hat{x}$) and in the direction of
average vorticity ($\hat{z}$), and Lees-Edwards boundary conditions in
the direction of the velocity-gradient ($\hat{y}$)~\cite{lee_edward}.
The volume fraction $\phi$, hence the box size is kept constant
throughout the simulation.

The particles interact via repulsive, finite-range contact forces
$\mathbf{f}$. A contact is made between particles $i$ and $j$, when
the shortest distance between the spines of the particles, $r_{ij}$ is
less than $D_{ij}$, where $D_{ij}=(D_{i}+D_{j})/2$.  The normal
$\mathbf{f}^n_{ij}$ and tangential $\mathbf{f}^t_{ij}$ components of
$\mathbf{f}_{ij}$, force on particle $i$ due to contact between $i$
and $j$, are calculated by a force-overlap relation as in a
linear-viscoelastic model,
\begin{align}
  \mathbf{f}^n_{ij}&=[k_n \delta_{ij} \mathbf{n}_{ij} - c_n\mathbf{v}^n_{ij}] ,\\
  \mathbf{f}^t_{ij}&=[-k_t \boldsymbol{\xi} ^t_{ij} - c_t\mathbf{v}^t_{ij}].
%\mathbf{f}^n_{ij}&=[k_n \Delta_{ij} \mathbf{n}_{ij} - C_n m_{eff}\mathbf{v}^n_{ij}] ,\\
%\mathbf{f}^t_{ij}&=[-k_t \mathbf{\xi} ^t_{ij} - C_t m_{eff}\mathbf{v}^t_{ij}].
\end{align}
Here, the normal direction $\mathbf{n}_{ij}$ is a unit vector pointing
from particle $j$ to $i$, $k_n$ and $k_t$ are spring constants, $c_n$
and $c_t$ are viscous damping constants,
%$m_{eff}=m_i m_j/(m_i+m_j)$ is effective mass,
$\boldsymbol{\xi} ^t_{ij}$ is elastic shear displacement pointing
tangential to the contacting particles, and $\mathbf{v}^n_{ij}$ and
$\mathbf{v}^t_{ij}$ are relative velocity comnonents in normal and
tangential direction. The relative velocities include translational
and angular components at the point of contact.  Particle $j$
experiences the same but opposite force $-\mathbf{f}_{ij}$. We have
set $k_n$ to unity, $k_t/k_n$ as 2/7, $c_t$ as zero and $c_n$ is such
that the coefficient of restitution equals 0.7.
All % ($c_n=0.16$). All
particles have the same mass set at 1.0.

Solid sliding friction is taken into account replacing $\mathbf{f}^t$
by
$\mathbf{f}^t=\mu(\frac{|\mathbf{f}^n|}{|\mathbf{f}^t|})\mathbf{f}^t$,
whenever the Coulomb inequality $|\mathbf{f}^t|>\mu |\mathbf{f}^t|$
holds. We have considered $\mu$=0.0, 0.5 and 1.0.

The motion of each particle is goverened by the collective force
gathered from all particles in contact.  The elastic and dissipative
forces give rise to torques on the particles, and rotation of particle
$i$ is dictated by
\begin{equation}
 I_i\dot{\mathbf{\omega}}_i=\sum_j \mathbf{a}_{ij} \times \mathbf{f}_{ij}
\end{equation}
where $I_i$ is the moment of inertia of particle $i$ and
$\mathbf{a}_{ij}$ is moment of arm from the center of particle $i$ to
the point of contact with particle $j$.

We integrate the equations of motion on a GPU using a velocity Verlet
algorithm for the translational degrees of freedom, and a
Richardson-like iteration for the rotational degrees of freedom, which
are represented as quaternions. Normalization of the quaternions is
ensured by rescaling at each time step.
%to include calculation of rotational
%motion~\cite{martys_mountain}.
The time-step of the integration is 0.01 and number of particles $N$ is
1000.

\section{Results}

\subsection{Rheology for aspect ratio $\mathrm{AR}=1.0$}

\begin{figure*}[t!]
  \includegraphics[width=0.3\textwidth]{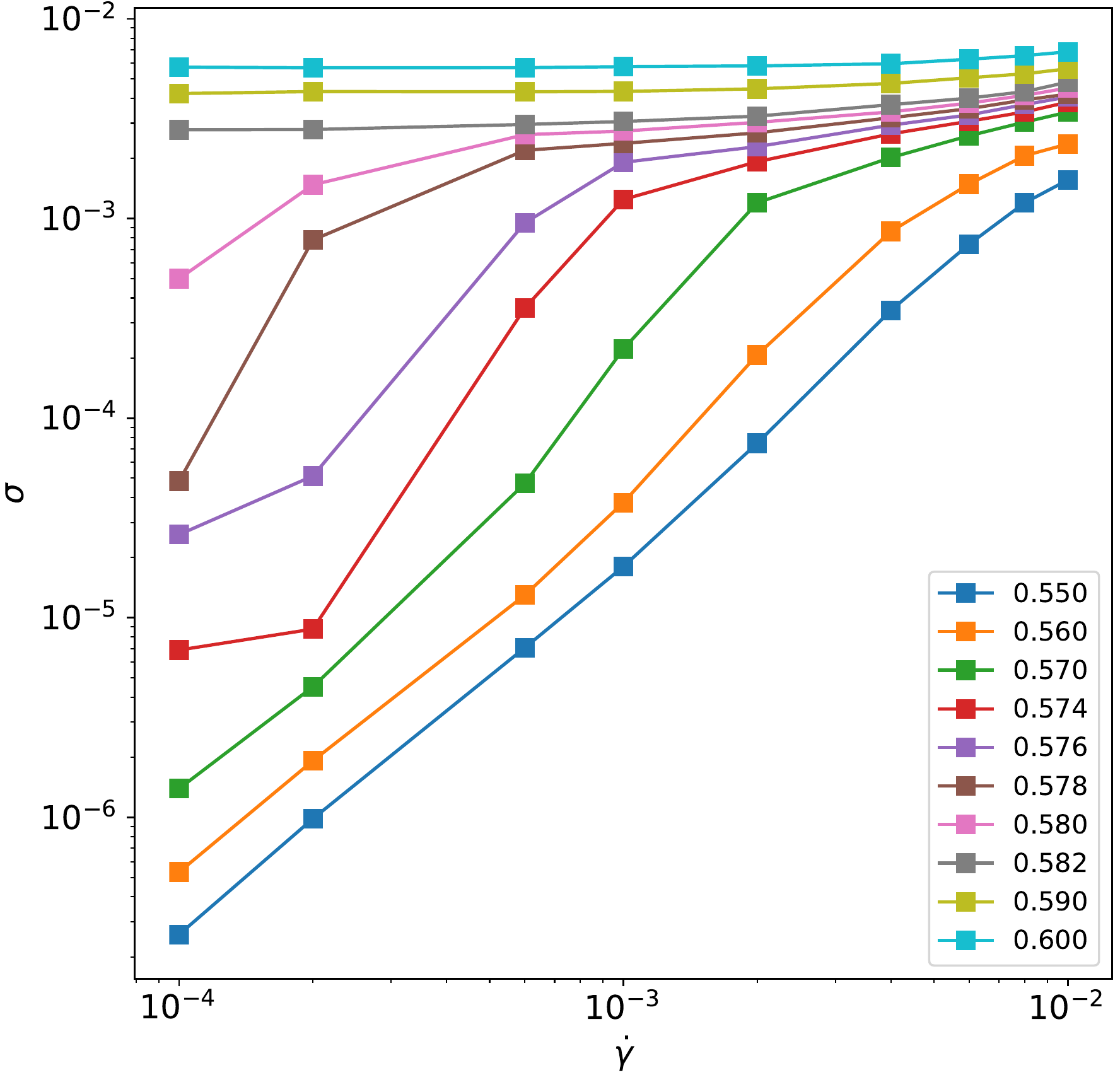}
  \includegraphics[width=0.3\textwidth]{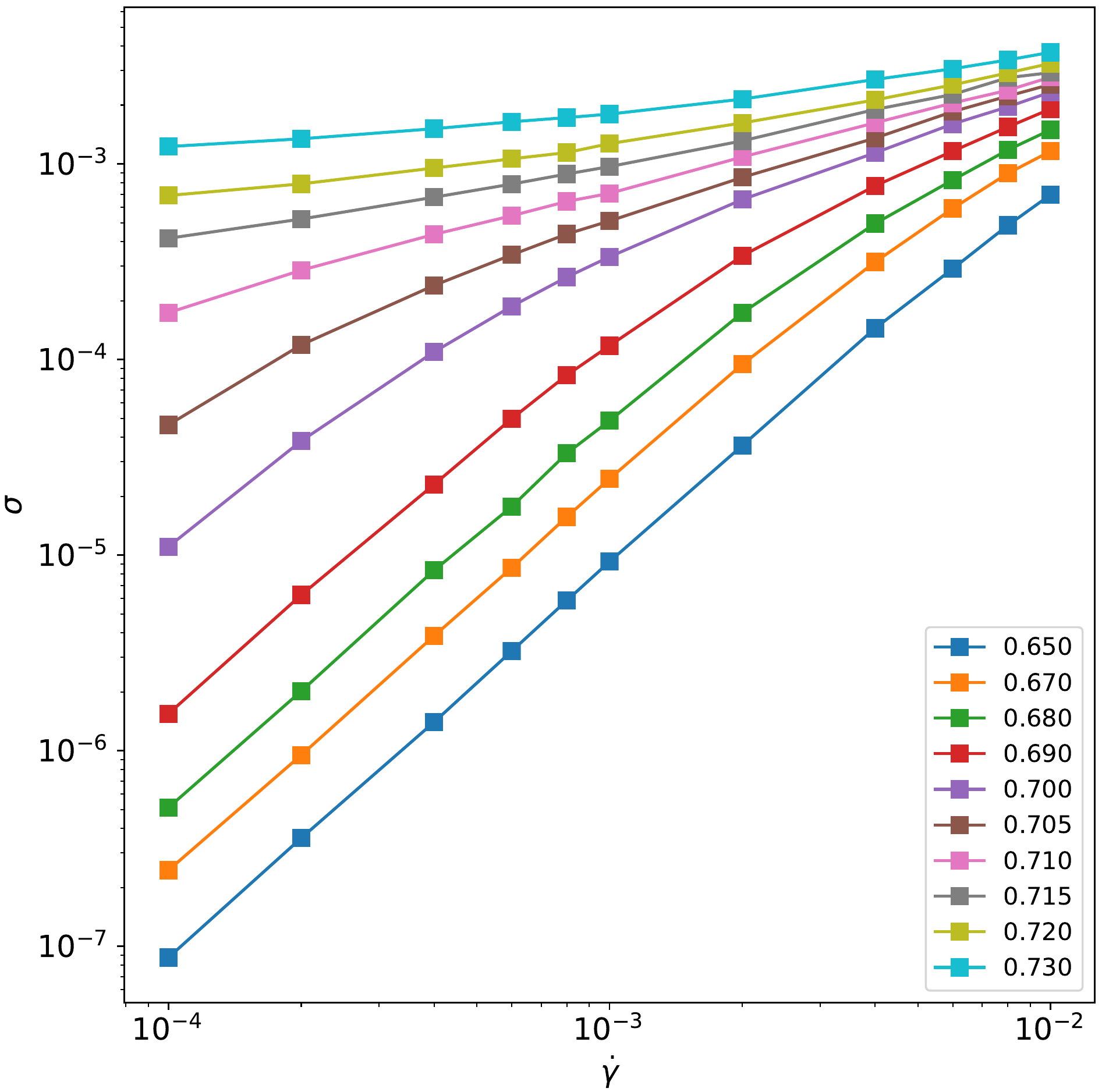}
  \includegraphics[width=0.3\textwidth]{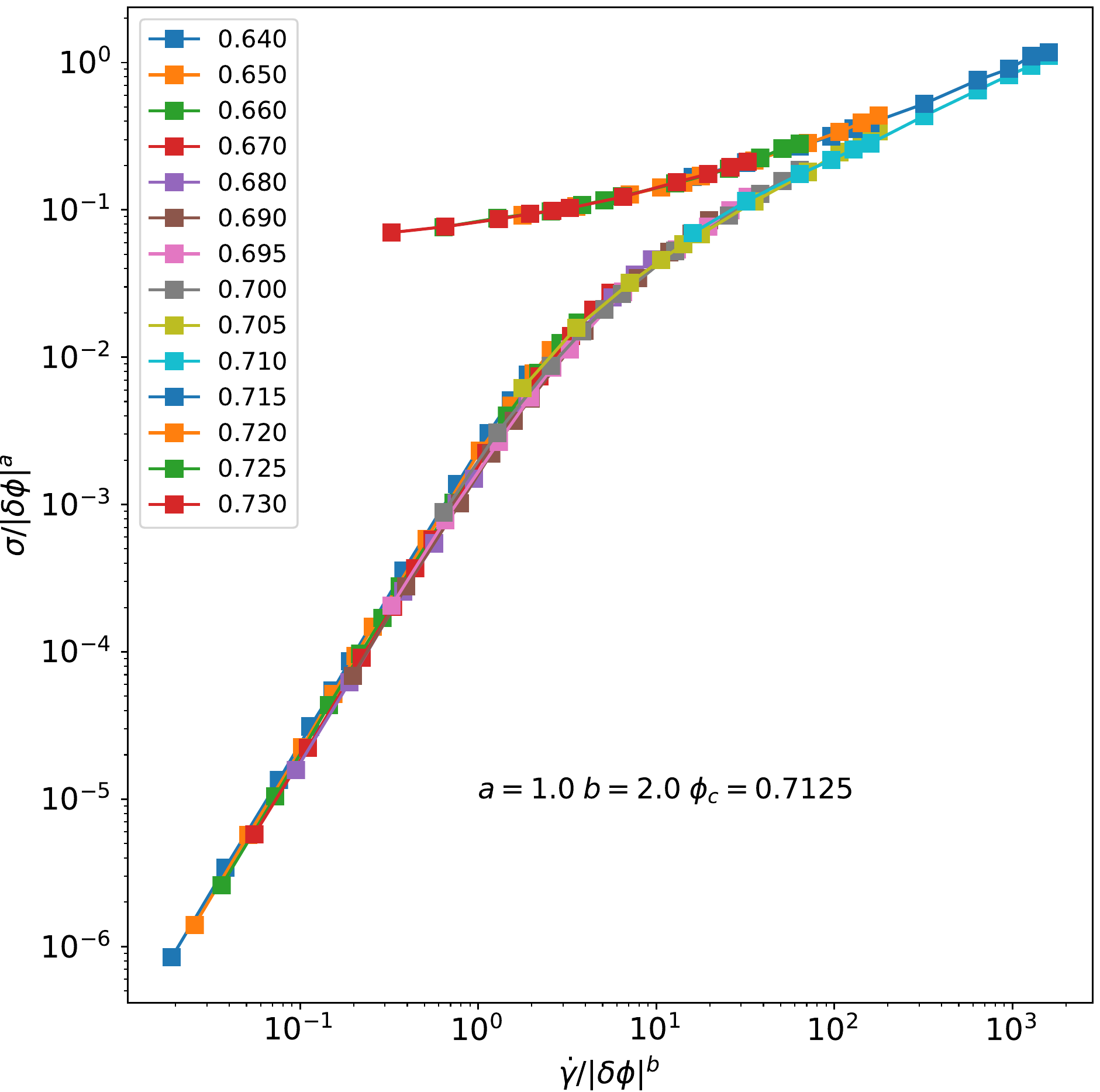}
  \caption{Stress $\sigma$ vs. strainrate $\dot\gamma$ for different
    volume fractions $\phi$ for spherocylinders of length
    $\mathrm{AR}=1.0$. Comparison of frictional (left) and frictionless
    (center) systems. (right) Scaling plot of the frictionless data,
    for details see text.}
  \label{fig:flowcurve}
\end{figure*}

We start the discussion with the rheological properties of
spherocylinders with a given aspect-ratio $\mathrm{AR}=1.0$. In
Fig.~\ref{fig:flowcurve} we present the flowcurves, i.e. the relation
between strainrate $\dot\gamma$ of the shear flow and the stress
$\sigma$ necessary to drive the flow. In the figure, we compare the
rheology of a frictional system ($\mu=1.0$, left) with that of a
frictionless system (i.e. $\mu=0$, center).

These flowcurves look rather similar to the much studied system of
spherical
particles~\cite{PhysRevE.93.030901,PhysRevE.89.050201,otsuki_hayakawa_2011}. The
frictional system has three distinctive regimes. At small
volume-fractions $\phi$ and at small strainrates $\dot\gamma$ the
stress shows Bagnold scaling, $\sigma\propto \dot\gamma^2$,
characteristic of inertial dynamics of hard particles. At high
volume-fractions, beyond a jamming threshold, the flow follows the
typical Herschel-Bulkley form (HB),
\begin{eqnarray}\label{eq:hb}
  \sigma = \sigma_y +c\dot\gamma^x\,,
\end{eqnarray}
with a yield stress $\sigma_y$ and
an HB exponent $x$. Delimited by these two asymptotic regimes we find
an additional shear-thickening regime (ST), which is a consequence of
the frictional interactions. ST implies a stronger increase of the
stress with strainrate than given by Bagnold scaling. In agreement
with previous work~\cite{PhysRevE.88.050201,PhysRevLett.111.218301,wyart_cates} it seems
that the onset of ST is not given by a characteristic strainrate but
by a stress $\sigma_0\approx 10^{-2}\sigma_y$. The physical origin of
this value is unclear at present.

The frictionless system does not have such a ST regime. Instead it
displays a continuous crossover-scenario from Bagnold to
shear-thinning behavior below a critical volume fraction $\phi_c$, and
from yield-stress to the same shear-thinning behavior above
$\phi_c$. Such a scenario corresponds to a continuous transition and
is well known from spherical-particle systems~\cite{olsson_teitel}. On
the other hand, the properties of the frictional system are
characteristic of a discontinuous transition like, for example, a
liquid-gas coexistence. For the continuous transition, we can use a
scaling Ansatz for the flowcurves
\begin{eqnarray}\label{eq:scaling_flowcurve}
\sigma(\phi,\dot\gamma) = |\delta\phi|^aF(\dot\gamma\tau)
\end{eqnarray}
with a time-scale $\tau=|\delta\phi|^{-b}$, $\delta\phi=\phi-\phi_c$,
and approximately determine the exponents $a$, $b$ as well as $\phi_c$
(see Fig.~\ref{fig:flowcurve} right). The obtained values imply,
e.g. for the viscosity
$\eta=\sigma/\dot\gamma^2\sim |\delta\phi|^{-\mu}$ with
$-\mu=a-2b\approx -3$. Likewise, the HB exponent defined in
Eq.~(\ref{eq:hb}) is given by $x=a/b\approx0.5$.  However, care must
be taken as the values of the exponents sensitively depend on the
value for $\phi_c$. Furthermore, our systems are rather small, and
finite-size effects are likely to strongly influence these values.

\subsection{Shear-induced alignment}

It is to be expected that the spherocylindrical particles align with
the flow. To quantify this effect we compute the average orientational
ordering tensor~\cite{orientation_tensor},
\begin{equation}
 T_{\mu\nu} =\Bigl \langle \frac{3}{2N} \sum_{i=1}^N \Bigl [\hat{\text l}_{i\mu} 
 \hat{\text l}_{i\nu}
 -\frac{\delta_{\mu\nu}}{3}] \Bigr \rangle
\end{equation}
where $\mathbf{\hat{\text l}}_{i}$ is the unit vector pointing along
the long axis of particle $i$. The quantity $T$ is a $3\times 3$
matrix, the largest eigenvalue of which is called the global nematic
order parameter $\mathbf{S2}$. The local nematic order parameter is
given by,
\begin{equation}\label{eq:local_align}
  s_{i}=\frac{3\hat{\text l}_{ix} \hat{\text l}_{ix}-1}{2}
\end{equation}
which is the $(x,x)^{th}$ entry of the local orientational tensor of
particle $i$, and represents a measure for the alignment with the flow
direction. Snaphshots of the system highlighting the local ordering
are given in Fig.~\ref{fig:snapshot_orient}.

\begin{figure}[h!]
  \includegraphics[width=0.23\textwidth]{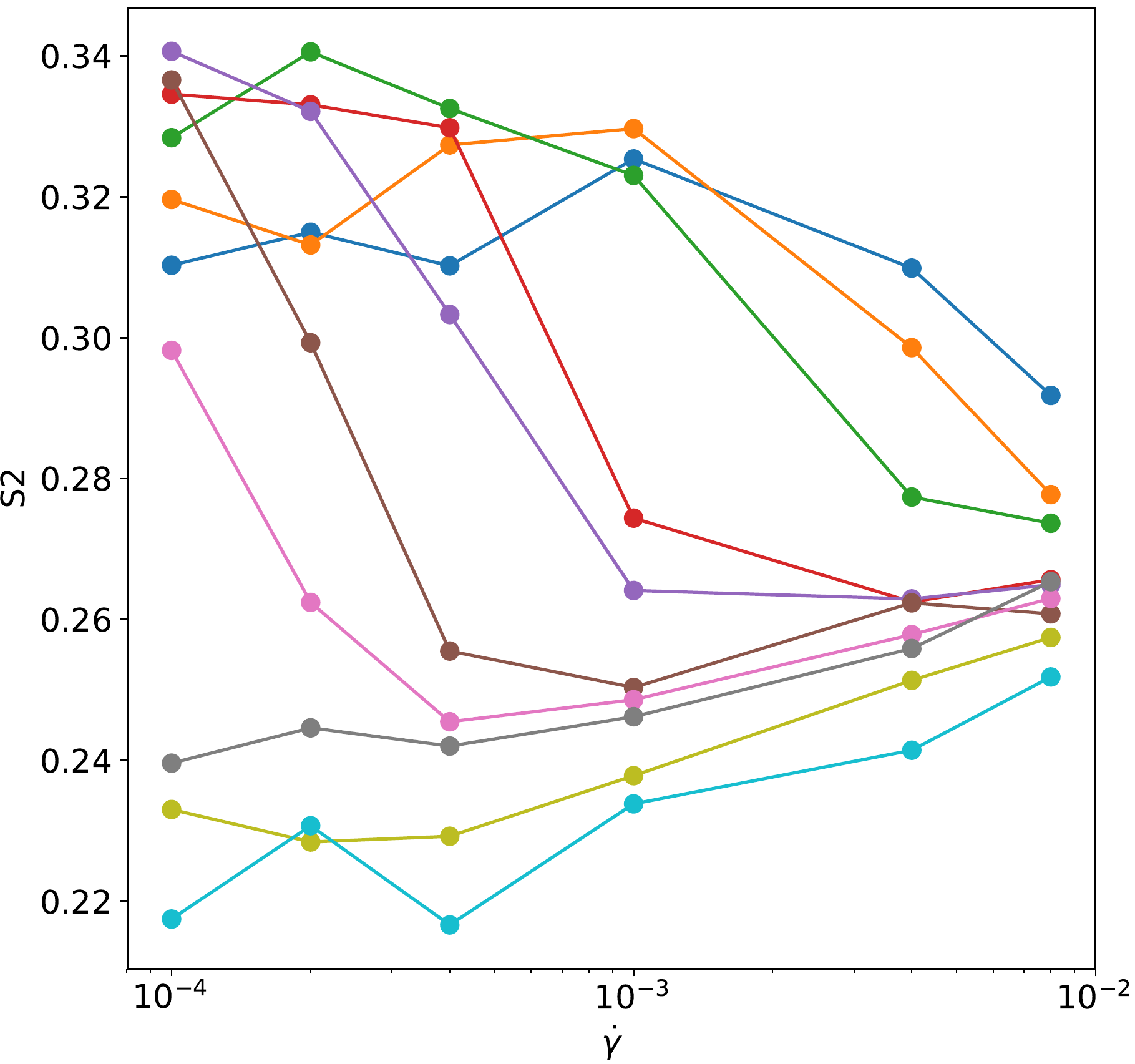}
  \includegraphics[width=0.225\textwidth]{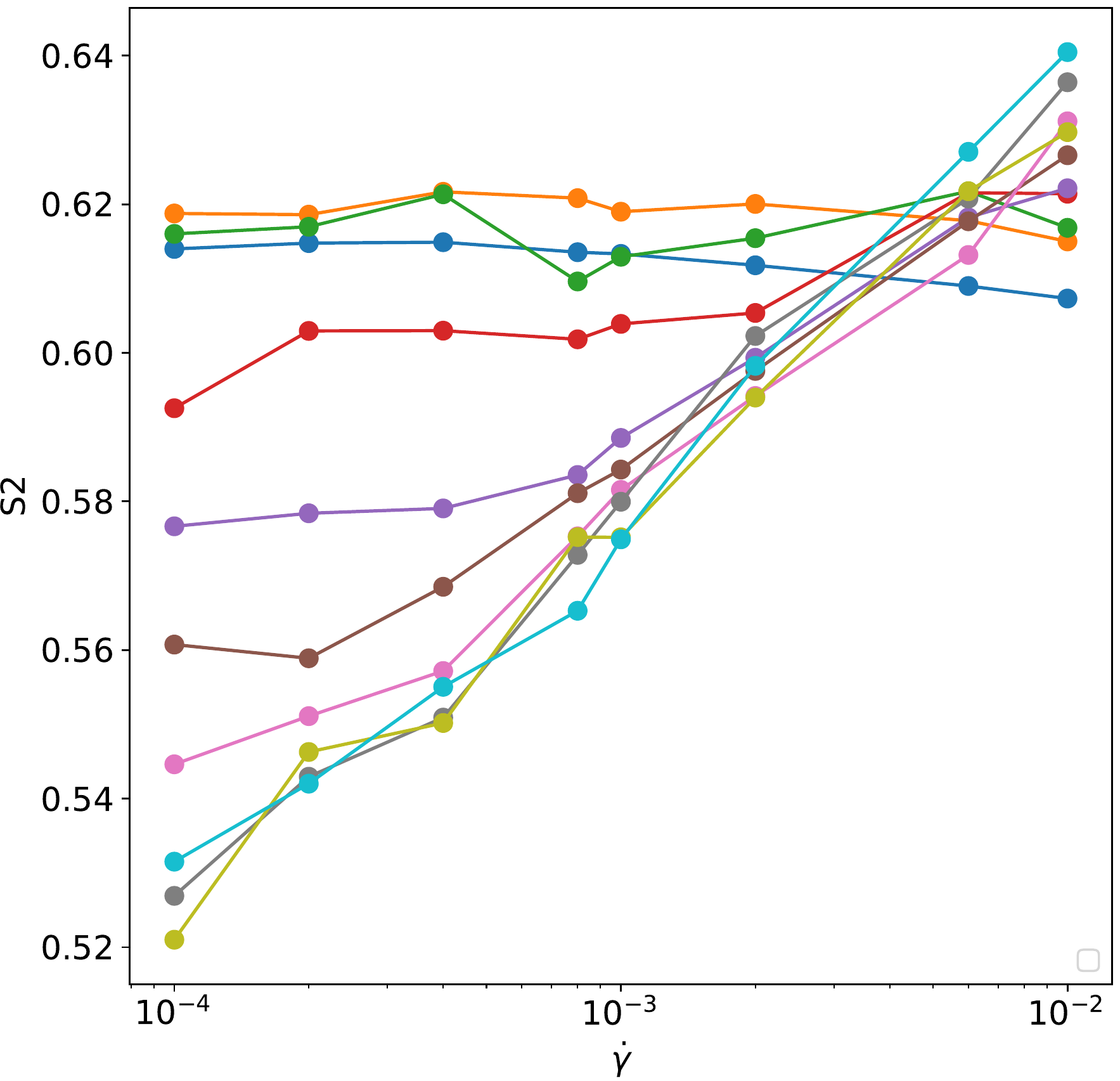}
  \caption{Orientational order parameter $\mathbf{S2}$ vs. strainrate
    $\dot\gamma$ for different volume fractions $\phi$ for
    spherocylinders of length $\mathrm{AR}=1.0$. $N=1000$ -- (left)
    frictional ($\mu=1.0$) and (right) frictionless system
    $\dot\gamma$.}
  \label{fig:nop}
\end{figure}

The global nematic order parameter is given in Fig.~\ref{fig:nop},
again comparing frictional and frictionless systems.
%Anticipating the specific role of the onset stress for ST $\sigma_0$
%we plot $\mathbf{S2}$ as a function of stress for the frictional
%system.
For the frictional system, in the fluid regime (small $\phi$ and
$\dot\gamma$) the nematic order is rather high, at least at the $\phi$
values close to jamming that we consider. In the HB regime (high
$\phi$), on the other hand, $\mathbf{S2}$ is small, possibly because
of plastic events leading to efficient randomization. In the ST regime
$\mathbf{S2}$ rapidly interpolates between the high value in the fluid
and the small value in HB. The strainrate at which this transition
happens corresponds to the onset of ST in Fig.~\ref{fig:flowcurve} and
represents the characteristic stress $\sigma_0$.

If we now turn to the frictionless system, then the first observation
is that the scale of nematic order is strongly enhanced as compared to
the frictional system. Apparently, friction suppresses or frustrates
flow alignment. The transition from fluid to jammed state is also
different than for frictional systems: as in the flow-curves it
features a continuous crossover scenario from the fluid or the HB
branch into a critical branch at a volume-fraction dependent
strainrate. The strainrate-independent value of S2 in the fluid branch
therefore continuously decreases upon approaching the jamming
limit. In contrast, the frictional scenario is discontinuous. Two
branches are connected via a rapid crossover at the onset stress for
ST. Similar data for frictionless ellipsoids in 3d has been presented
in Ref. ~\cite{2018teitel}.

\subsection{Distribution of alignment angles}

For ellipsoidal and cylindrical particles with small friction the
typical angle between the ellipsoid major axis and the flow has been
shown to be
non-zero~\cite{campbell_ellipse,guo_wassgren2012,guo_wassgren_2013}. Similar
results have been obtained for dumbbell-shaped
particles~\cite{reddy_2009} and in experiments of shear flows of
elongated particles in a cylindrical split bottom shear
cell~\cite{PhysRevLett.108.228302}.
%but the typical angles are somewhat smaller.

\begin{figure}[h!]
 \includegraphics[width=0.4\textwidth]{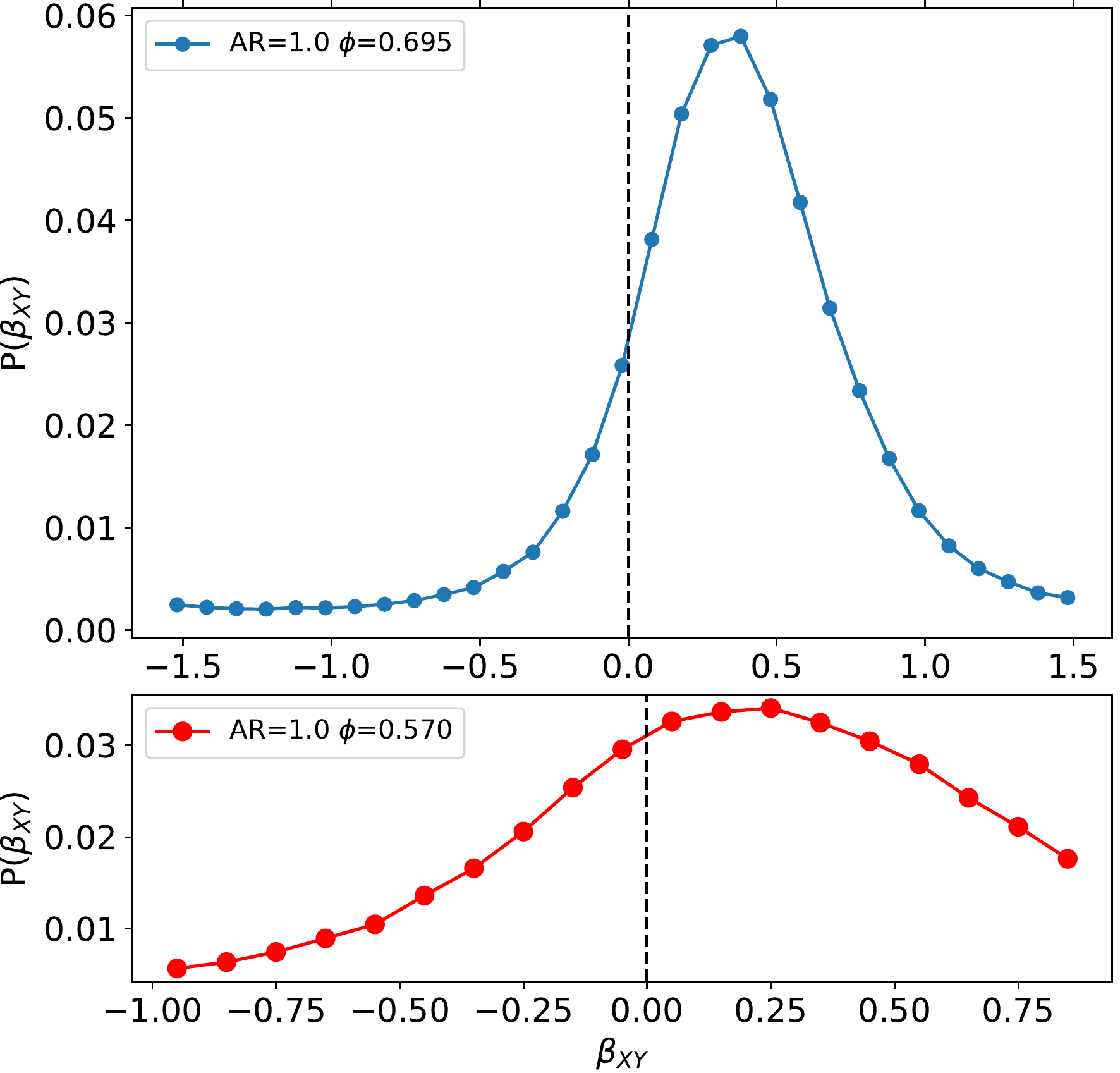}
 \caption{Probability distribution $P(\beta_{\rm xy})$ of angles
   $\beta_{\rm xy}$ with flow direction (angle between $x$-axis and
   particle projection into shear-gradient (xy) plane) for aspect
   ratio $\mathrm{AR}=1.0$, (top) frictionless and (bottom) frictional
   flows; strainrate $\dot\gamma=10^{-4}$ and volume fraction chosen
   to be in the fluid regime.}
  \label{fig:angle}
\end{figure}

We display the distribution of orientation angles in the fluid regime
in Fig.~\ref{fig:angle}. We define $\beta_{\rm xy}$ as the angle
between the flow direction and the projection of the particle
orientation into the shear-gradient (xy) plane. For the frictionless
system we observe a clear peak at a finite angle which, for the
frictional system, is strongly degraded.

The origin of the peak can be explained with the thermal tumbling
dynamics of a single particle, which spends a rather long time at
small positive
angles~\cite{hinch72:_brown,lee09:_tumbl_dynam_rod_semi_polym,PhysRevLett.96.038304}. A
negative angle, on the other hand, which may be achieved by a thermal
fluctuation, is unstable and immediately leads to full rotation of the
particle back into a stable positive angle. The smaller the
temperature (the larger the Peclet number) the more pronounced is this
effect. In our system, the temperature may be due to the interaction
with the surrounding particles. The frictional system then providing
higher effective temperature as the frictionless system. A similar
trend is observed when going into the plastic flow HB regime (not
shown). There the peak height is reduced in agreement with the
presumed higher randomization due to plastic events as mentioned
above.

% and have peaks at 39$^\circ$, 34$^\circ$, and 27$^\circ$ for aspect
% ratio 1.25, 1.5, and 2.0, whereas angle between flow direction and the
% projection of major axis on horizontal plane ($\beta_{XZ}$) is peaked
% around zero~\cite{campbell_ellipse}.  Location of peak is unchanged
% with volume fraction but the peaks gets narrower with increase in
% aspect ratio and volume fraction.  With increase in frictional forces,
% the same angular preferences exist with flatter distributions and
% broader peaks, particularly in the $\beta$.

% Cylindrical and glued-spheres frictionless particles have peaks at
% small and positive $\mathrm{AR}$ in the range 0 and
% 10$^\circ$~\cite{guo_wassgren2012,guo_wassgren_2013}. This is also
% observed for dumbbell-shaped particles~\cite{reddy_2009} and in
% physical experiments of shear flows of elongated particles in a
% cylindrical split bottom shear cell, where $\alpha$ becomes 6$^\circ$
% or 7$^\circ$~\cite{PhysRevLett.108.228302}.

\subsection{Contacts for short spherocylinders, $\mathrm{AR}=0.1$}

Two recent publications \cite{somfai,2018teitel} describe an
interesting phenomenon relating to the number of contacts between
(frictionless) spherocylinders/ellipsoids during flow. Apparently,
nearly-spherical particles tend to establish an excessive number of
contacts at their sides. Notably, the area of the side is rather small
as compared to the spherical cap region.

\begin{figure}[h!]
  \includegraphics[width=0.4\textwidth]{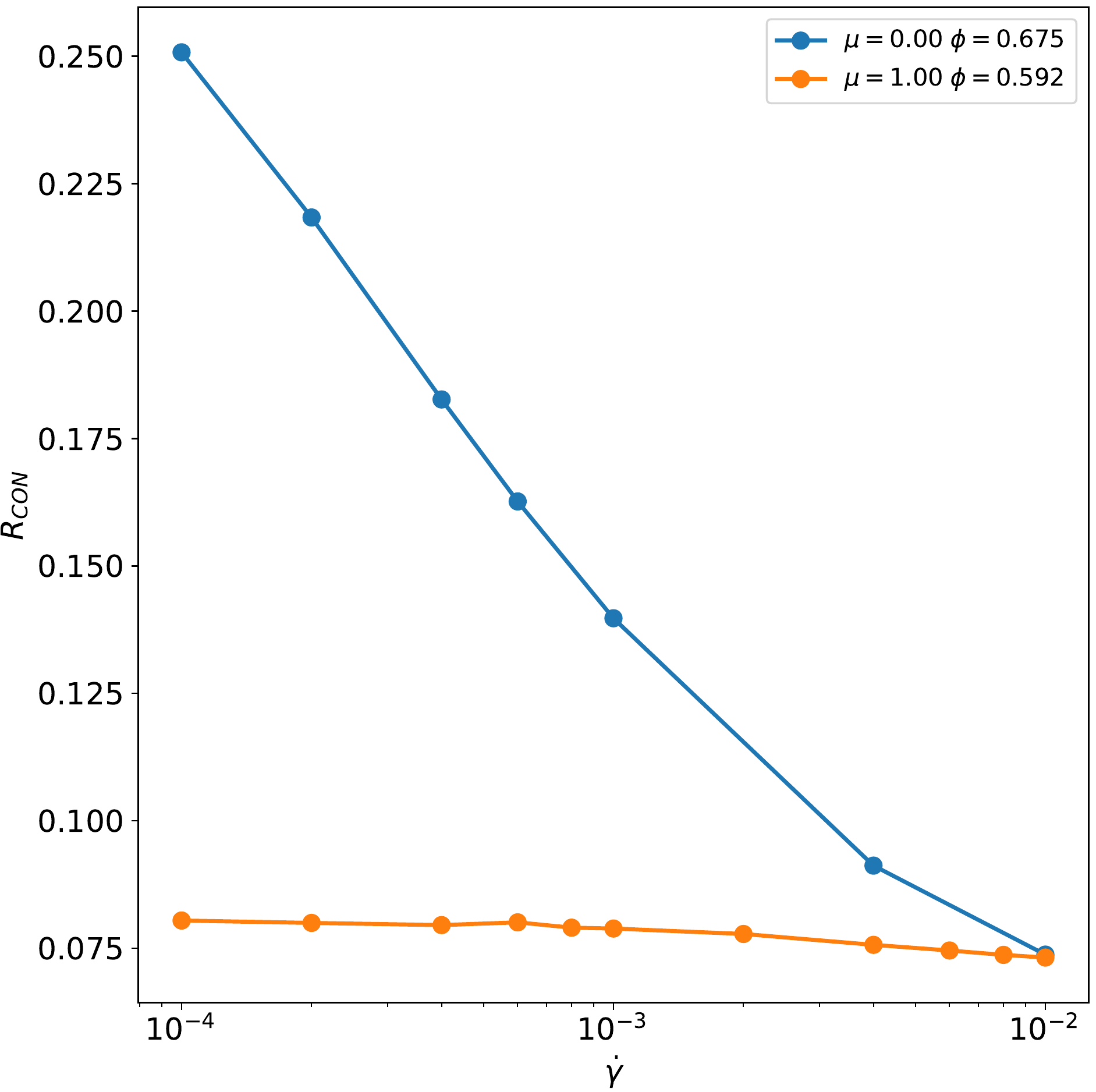}
  \caption{Ratio of side-to-end contacts vs. strainrate $\dot\gamma$
    at the jamming transition $\phi=\phi_c$ and
    $\mathrm{AR}=0.1$. Comparison of frictional and frictionless
    system.}
  \label{fig:contacts}
\end{figure}

In Fig.~\ref{fig:contacts} we plot the ratio $R_{\rm con}$ of
side-to-end contacts comparing the frictionless with the frictional
system.  From the range of values it is readily observed that the
number of side contacts in the frictionless simulations are increased
as compared to the frictional case.  The fraction of side contacts
rises to about $25\%$ without friction and $~8\%$ with friction. The
latter value can be understood from the ratio $r$ of surface areas on
the side ($\pi DL$) and on the caps ($\pi D^2$), giving
$r=\mathrm{AR}=0.1$. Thus, the number of end contacts in the
frictional scenario is to be considered normal, while that in the
frictionless scenario is markedly enhanced.

We also set up additional simulations for the case of isotropic
jamming without shear. To this end we run a compression-decompression
cycle with $\phi_{\rm min}=0.64$ and $\phi_{\rm max}=0.74$. From the
decompression branch the jamming transition can be located at
$\phi_c\approx 0.68\ldots 0.685$ (the error stems from the step-size).
At jamming we find roughly $5$ end contacts per particle, as well as
$1.6$ side contacts. For all volume fractions above $\phi_c$ the ratio
of side-to-end contacts is constantly about $32\%$ and therefore even
higher than in our shearing simulations. With the strong increase
towards lower strainrates, the latter high value may eventually be
reached, however.

This strong increase visible in Fig.~\ref{fig:contacts} is indeed
remarkable, as it indicates the presence of a long (but necessarily
finite) time scale, at which the contact numbers cross over into their
zero-strainrate limit. This time-scale cannot be the same as the
$\tau$ extracted from the scaling of the flowcurve,
Eq.~(\ref{eq:scaling_flowcurve}), as that diverges at the jamming
transition. One may speculate that this time-scale is connected to
small-scale rotational motion. In packings of nearly-spherical
ellipsoids rotational modes occur at the lower end of the frequency
spectrum~\cite{zeravcic2009EPL,PhysRevLett.102.255501}, thus possibly
leading to a small time-scale visible under shear. Longer simulations
at smaller strainrates are necessary to answer this question in
detail.

\subsection{Shear thickening}

If we compare data from different aspect ratios $\mathrm{AR}$ we
expect the amount of shear alignment to increase with increasing
$\mathrm{AR}$ \cite{PhysRevLett.108.228302,2018teitel}. This may
already be apparent from the snapshots in
Fig.~\ref{fig:snapshot_orient}. What is the consequence for the
rheological properties? At first sight stronger alignment should make
flow easier. This is what we have observed in the fluid regime in
Fig.~\ref{fig:nop}, increasing alignment corresponds to decreasing
stress. Note, however, that the opposite is true for the HB regime,
where plastic events supposedly counteract alignment and lead to small
$\mathbf{S2}$.

\begin{figure}[h!]
  \includegraphics[width=0.4\textwidth]{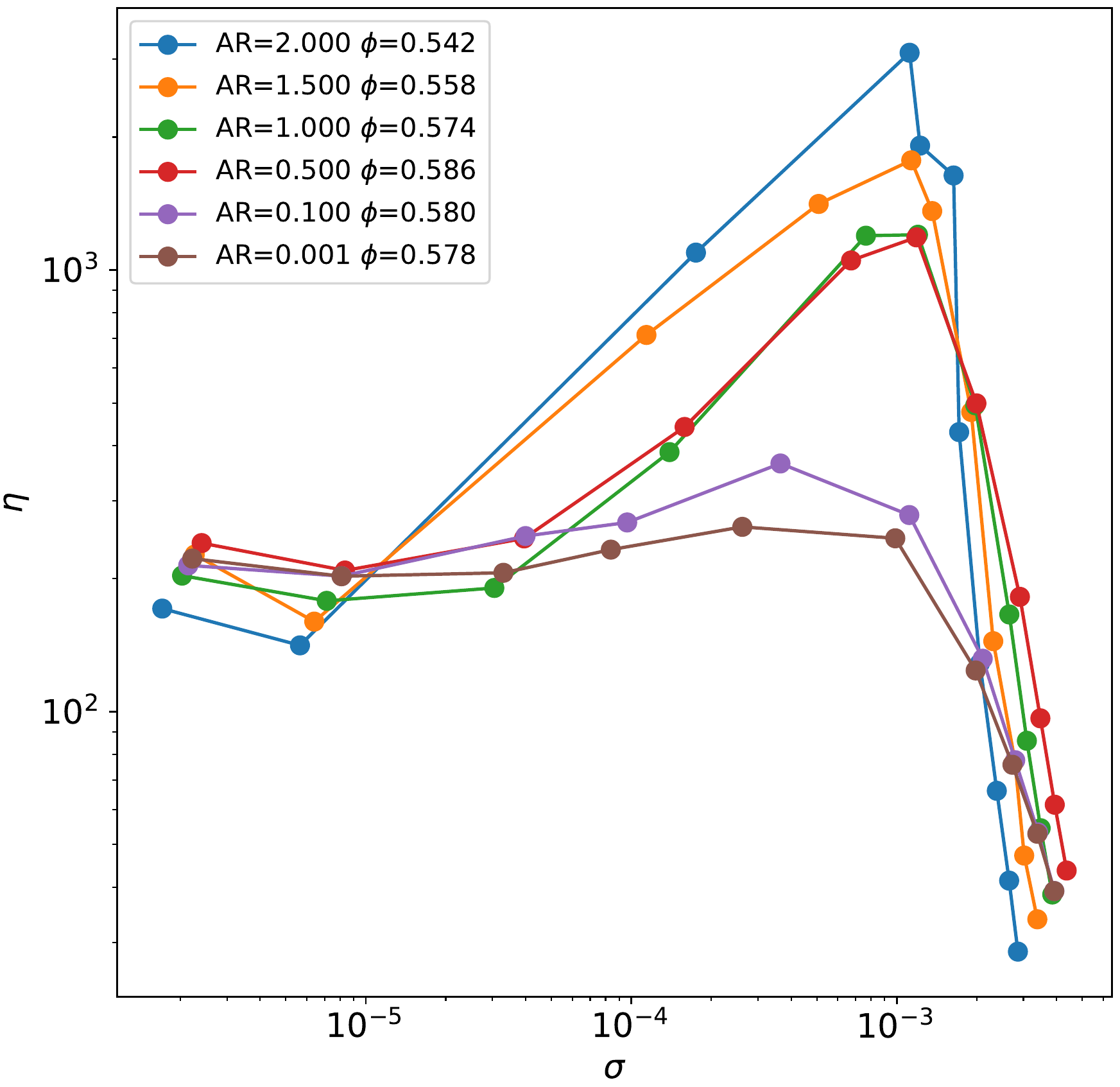}
  \caption{Viscosity $\eta=\sigma/\dot\gamma^2$ vs. stress
    $\sigma$ for different aspect ratios $\mathrm{AR}=0.001\ldots 2.0$. The
    volume fraction for each $\mathrm{AR}$ is chosen such that the
    zero-stress viscosity is approximately the same for each.
    %(right) Stress $\sigma$ vs. strainrate $\dot\gamma$ for different
    %aspect ratios $\mathrm{AR}$. The volume fraction $\phi$ for each
    %$\mathrm{AR}$ is chosen such that the $\phi/\phi_J$ is approximately
    %the same for each.
  }
  \label{fig:viscosity_ar}
\end{figure}

The rheology for aspect ratios $\mathrm{AR}\neq 1$ is not much
different from Fig.~\ref{fig:flowcurve}. The relevant range of volume
fractions changes with the aspect ratio. We chose to use the
viscosity in the fluid regime to compare these data. In
Fig.~\ref{fig:viscosity_ar} we therefore compare effective viscosities
$\eta\equiv\sigma/\dot\gamma^2$ for different $\mathrm{AR}$ but with
similar low-strainrate limit $\eta(\dot\gamma\to 0)=\eta_0$. What is
most apparent for these data is the effect of aspect ratio on
thickening. While short particles display only a mild viscosity
increase in the ST regime, long particles have a much stronger
increase -- all starting from the same small-strainrate viscosity. At
the same time the ST regime spans to smaller stresses. Apparently, the
onset stress for ST decreases with the length of the particles. While
there is no general understanding of the parameter dependencies of
this onset stress, a negative correlation with particle length at
least is plausible, when one considers the effects of particle
rotations. In Ref.~\cite{maiti_claus} we have argued that increasing
stress leads to the build-up of rotationally frustrated structures
that may bear higher loads. As to the anisotropic shape of longer
particles, such an effect is likely to be reinforced. Thus, smaller
stresses are sufficient to lead to the same ``strength'' of ST.

In this context we also want to mention the two experiments
\cite{PhysRevE.84.031408,egres05}, which deal with the rheology of
cylindrical particles in the range $\mathrm{AR}=2\ldots 9$. While in
Ref. \cite{PhysRevE.84.031408} the onset stress is masked by a
shear-thinning regime at small stress, Ref.\cite{egres05} seem to find
the same onset stress for the different particle lengths (which are
longer than ours).

%yield stress is approximately the same for all lengths: maybe due to
%the fact that yield stress is dominated by number of particle
%contacts, specific orientation may not be as important

%the bigger the gap between onset and yield stress, the more pronounced
%the effect of ST and the higher the effective viscosity.

\subsection{Aspect-ratio dependence of $\phi_C$ }

\begin{figure}[h!]
  \includegraphics[width=0.4\textwidth]{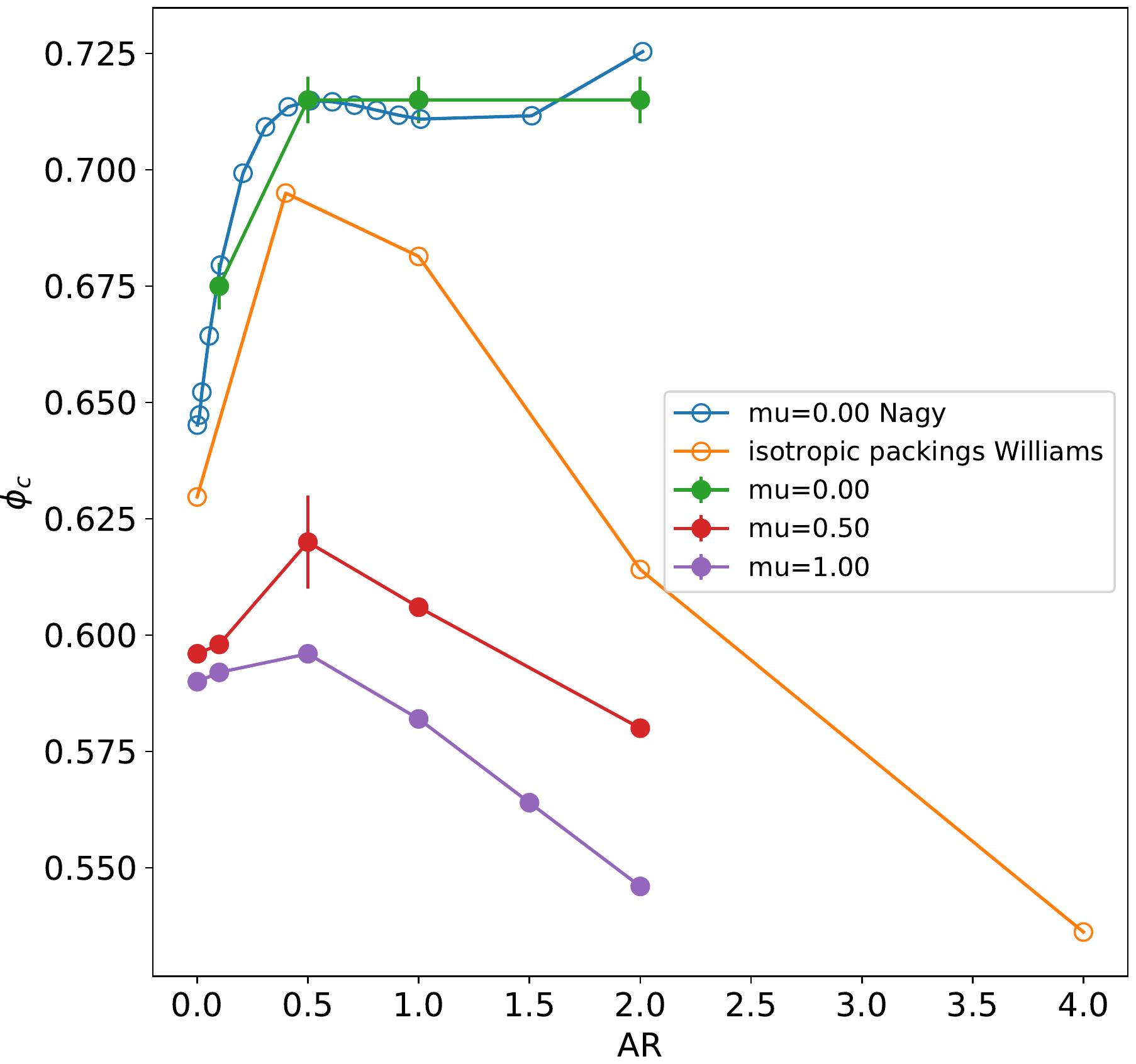}
  \caption{Jamming density $\phi_c$ vs. aspect-ratio $\mathrm{AR}$ for
    different friction coefficients $\mu$. Comparison with data taken
    from literature for $\mu=0$ (Nagy {\it et al.} \cite{somfai}) and
    for isotropic packings Williams {\it et
      al.}~\cite{PhysRevE.67.051301}. }
  \label{fig:jam_density}
\end{figure}

It is well known~\cite{PhysRevE.67.051301} that the jamming density
$\phi_c$ of isotropic packings of spherocylinders first increases with
$\mathrm{AR}$, then presents a maximum at $\mathrm{AR}\approx0.5$
before declining asymptotically as $\phi_c\propto \mathrm{AR}^{-1}$
(see Fig.~\ref{fig:jam_density}, data marked as ``Williams'').

The latter can be understood with an excluded volume
argument~\cite{F29868200317,phi96} that relates the excluded volume of
a section $L_e$ of a long cylinder $v_{\rm excl}\sim (\pi/2) L_e^2D$
to the available volume $v_{\rm avail}\sim 1/\rho z$, where $\rho$ is
the number density of cylinders and $z=L/L_e$ is the number of
sections of length $L_e$ per particle. In terms of volume fraction
$\phi\sim \rho (\pi/4) LD^2$ one arrives at $\phi \sim z/2\mathrm{AR}$. The
number of sections $z$ is equal to the number of contacts, which --at
jamming-- should be $z\approx 8$ for long frictionless cylinders. The
jamming volume fraction thus is given by
$\phi_c\approx 4/\mathrm{AR}$~\cite{rod05}. We do not probe this
asymptotic regime here, but remain in the range of the maximum.

As can be seen in the figure, frictionless particles in isotropic
packings (Williams {\it et al.}~\cite{PhysRevE.67.051301}) display a
rather strong increase of the jamming density when increasing the
length of the particles from the sphere limit $\mathrm{AR}=0$. It has
been argued~\cite{PhysRevE.75.051304} that nearly-spherical particles
can make more efficient use of the available space than spheres,
because orientations may be optimized within a given local
surrounding. In fact, the slope of an initial linear increase in
$\phi_c(\mathrm{AR})$ can be well approximated by satisfying
torque-balance constraints for each particle, given the local
surrounding of the equivalent sphere
packing~\cite{PhysRevE.75.051304}. This sets an ``optimal''
orientation of the particle which allows for calculating the jamming
density. On the other hand, chosing random orientations for the
particles one would obtain a vanishing slope, i.e. insensitivity of
$\phi_c$ towards aspect-ratio~\cite{PhysRevE.75.051304}.

Under shear the local surrounding is constantly changing, thus
additional rotations are necessary to fulfil the torque-balance
constraints. Recent pressure-controlled shear simulations with
frictionless particles~\cite{somfai} indicate that the strong increase
for small $\mathrm{AR}$ is preserved, albeit at slightly elevated
volume fractions $\phi$ (see Fig.~\ref{fig:jam_density} data marked as
``Nagy''). We confirm these data within our framework of
$\phi$-controlled simulations. Note, that a similar difference between
the jamming density in isotropic systems and that under shear is well
known from systems of spherical
particles~\cite{ohernPRL2002,PhysRevLett.102.218303}. At larger
$\mathrm{AR}$ the jamming density does not decrease as in the
isotropic packings. As discussed above these systems display
alignment. This induces correlations between particles such that the
excluded volume argument is invalidated~\cite{phi96}.

With friction, the relevant volume fractions are strongly reduced,
which is to be expected. What may be surprising, however, is that the
non-monotonic behavior is much less pronounced and the initial slope
is strongly decreased. At large enough friction coefficient $\mu$ the
maximum even seems to disappear. A similar effect is observed in a
two-dimensional flow of ellipses in Ref.~\cite{trulsson_2018}. Given
the role of particle rotations for the increasing $\phi_c$ for small
$\mathrm{AR}$ as discussed above, it is tempting to ascribe this
effect to the tangential frictional forces acting parallel to the
surface of the particle. These very effectively resist particle
rotation. Thus, additional packing optimization due to particle
rotation, which seems very prominent in nearly-spherical frictionless
particles, is absent in the presence of friction. As remarked in
Ref.~\cite{PhysRevE.75.051304}, insensitivity of the jamming density
towards aspect-ratio could result from randomly orienting the
particles irrespective of their local surrounding.

%dynamically, might take a long time to equilibrate along the
%associated direction in phase space\cite{PhysRevE.75.051304}, soft mode\cite{zeravcic2009EPL}

 %\begin{figure*}[h!]
 %\includegraphics[width=0.02\textwidth]{color_map.png}
 %   \includegraphics[width=0.29\textwidth]{color_angle_betn_orientaion_xy_proj_AR_0.001_phi_0.582_1.0e_04.png}
 %    \includegraphics[width=0.29\textwidth]{color_angle_betn_orientaion_xy_proj_AR_1.0_phi_0.576_1.0e_04.png}
 %    \includegraphics[width=0.29\textwidth]{color_angle_betn_orientaion_xy_proj_AR_2.0_phi_0.542_1.0e_04.png}
 %     \caption{Snapshots of the system at steady state. The particles are colored according to its orientation
 %  with respect to the XY-plane. Snapshots of systems with strainrate $1.0e-04$, with particles having (a) $AR$=0.001, 
 %  $\phi$=0.582, (b) $AR$=1.0, $\phi$=0.576 and (c) $AR$=2.0, $\phi$=0.542}
 %  \label{fig:snapshot_orient}
 %\end{figure*}

\section{Summary}

We have studied, by molecular dynamics simulations, the shear rheology
of dense packings of soft spherocylinders, highlighting the
differences between frictional and frictionless systems. We
concentrate on the regime of short spherocylinders with aspect ratios
$\mathrm{AR}=0.001\ldots 2.0$, close to the sphere limit. The
flowcurves we obtain are rather similar to those of spherical
particles. In particular, we find a shear-thickening regime (ST) in
the frictional system, just as with spheres. The frictionless system,
on the other hand, only shows shear-thinning behavior and no shear
thickening. When comparing different particle aspect ratios we find
that longer particles have a smaller onset stress for ST. Starting
with the same small-stress viscosity, the maximal increase in
viscosity can thus be modulated by particle shape. Friction also
affects the strength of shear-induced alignment, with frictional
systems showing a much smaller alignment than frictionless
systems. Furthermore, frictionless systems display a rather large
percentage of side-contacts between spherocylinders. It is much
larger than in frictional systems, where the percentage roughly
reflects the surface area of the side of the spherocylinder relative
to its total area.

In order to rationalize these phenomena we highlight the importance of
particle rotations, which follows from the nature of frictional forces
to act tangential to particle surfaces. Frictional torques from the
fluctuating dense environment may act as additional noise on the
particle orientation, thus reducing overall alignment. We have
discussed this question in terms of the probability distribution of
orientation angle of the particles. This is much flatter for
frictional particles than for frictionless particles. In addition
frictional forces may be very effective in suppressing small-scale
rotations that could optimize local packing arrangements. This could
also be responsible for the observation that in frictional systems the
jamming threshold is quite insensitive to aspect ratio, in contrast to
what is known from frictionless systems.

%With nearly-spherical particles, modes connected to
%rotations are much softer than translational modes. The associated
%long time-scale is visible in the frictionless system in the number of
%particle-end contacts, but not in the frictional system, where these
%modes are stabilized by the frictional forces.

\begin{acknowledgments}
  We acknowledge discussions with L. Pournin. %financial support by ...
\end{acknowledgments}

%merlin.mbs apsrev4-1.bst 2010-07-25 4.21a (PWD, AO, DPC) hacked
%Control: key (0)
%Control: author (8) initials jnrlst
%Control: editor formatted (1) identically to author
%Control: production of article title (-1) disabled
%Control: page (0) single
%Control: year (1) truncated
%Control: production of eprint (0) enabled
%

%\bibliographystyle{unsrt}
%\bibliographystyle{spphys}
%\bibliography{cite-paper}
\end{document}